\documentclass[]{spie}  

\usepackage{amsmath,amsfonts,amssymb}
\usepackage{graphicx}
\usepackage[colorlinks=true, allcolors=blue]{hyperref}

\title{From Assembly to the Complete Integration and Verification of the SOXS Common Path.}

\author[a]{Kalyan Kumar Radhakrishnan Santhakumari}
\author[a]{Federico Battaini}
\author[a]{Riccardo Claudi}
\author[a,u]{Alessandra Slemer}
\author[b]{F. Biondi}
\author[c]{M. Munari}
\author[c]{R. Z. Sanchez}
\author[d]{M. Aliverti}
\author[d]{L. Oggioni}
\author[e]{M. Colapietro}
\author[a]{D. Ricci}
\author[a]{L. Lessio}
\author[a]{M. Dima}
\author[a]{L. Marafatto}
\author[a]{J. Farinato}
\author[d]{S. Campana}
\author[e]{P. Schipani}
\author[e]{S. D'Orsi}
\author[a]{B. Salasnich}
\author[a]{A. Baruffolo}
\author[f]{S. Ben-Ami}
\author[e]{G. Capasso}
\author[g]{R. Cosentino}
\author[h]{F. D'Alessio}
\author[d]{P. D'Avanzo}
\author[f]{O. Hershko}
\author[i]{H. Kuncarayakti}
\author[d]{M. Landoni}
\author[j]{G. Pignata}
\author[k]{A. Rubin}
\author[l]{S. Scuderi}
\author[h]{F. Vitali}
\author[m]{D. Young}
\author[p]{J. Achrén}
\author[n]{José Antonio Araiza-Durán}
\author[r]{I. Arcavi}
\author[n]{Anna Brucalassi}
\author[f]{R. Bruch}
\author[a]{Enrico Cappellaro}
\author[d]{M. Della Valle}
\author[b]{R. Di Benedetto}
\author[f]{A. Gal-Yam}
\author[d]{Matteo Genoni}
\author[g]{M. Hernandez}
\author[r]{J. Kotilainen}
\author[s]{G. Li Causi}
\author[c]{L. Marty}
\author[i]{S. Mattila}
\author[f]{Michael Rappaport}
\author[c]{M. Riva}
\author[o]{S. Smartt}
\author[t]{M. Stritzingerv}
\author[g]{H. Venturae}

\affil[a]{INAF - Osservatorio Astronomico di Padova, Vicolo dell’Osservatorio 5, I-35122 Padova, Italy}
\affil[b]{Max-Planck-Institut f\"ur Extraterrestrische Physik, Giessenbachstr. 1, D-85748 Garching}
\affil[c]{INAF - Osservatorio Astrofisico di Catania, Via S. Sofia 78, I-95123 Catania}
\affil[d]{INAF - Osservatorio Astronomico di Brera, Via Bianchi 46, I-23807 Merate (LC)}
\affil[e]{INAF - Osservatorio Astronomico di Capodimonte, Salita Moiariello 16, I-80131 Napoli}
\affil[f]{Weizmann Institute of Science, Rehovot, Israel}
\affil[g]{INAF - Fundación Galileo Galilei, Breña Baja, Spain}
\affil[h]{INAF - Osservatorio Astronomico di Roma, Roma, Italy}
\affil[i]{Tuorla Observatory, Department of Physics and Astronomy, University of Turku, Finland}
\affil[j]{Universidad Andres Bello, Avda. Republica 252, Santiago, Chile}
\affil[k]{European Southern Observatory, Karl Schwarzschild Strasse 2, D-85748, Garching}
\affil[l]{Universidad Andres Bello, Santiago, Chile}
\affil[m]{European Southern Observatory, Garching, Germany}
\affil[n]{INAF - Istituto di Astrofisica Spaziale e Fisica Cosmica, Milano, Italy}
\affil[o]{Queen's University Belfast, School of Mathematics and Physics, Belfast, UK}
\affil[p]{Incident Angle Oy, Turku, Finland}
\affil[q]{INAF-Osservatorio Astrofisico di Arcetri, Firenze, Italy}
\affil[r]{Tel Aviv University, Tel Aviv, Israel}
\affil[s]{INAF - Istituto di Astrofisica e Planetologia Spaziali, Roma, Italy}
\affil[t]{Aarhus University, Aarhus, Denmark}
\affil[u]{Officina Stellare S.p.A., Sarcedo, Italy}

\authorinfo{Author E-mail: kalyan.radhakrishnan@inaf.it, Telephone: +39-049-8293-447}

\begin{document} 
\maketitle

\begin{abstract}
The Son Of X-Shooter (SOXS) is a single object spectrograph offering simultaneous spectral coverage in UV-VIS (350-850~$nm$) and NIR (800-2000~$nm$) wavelength regimes with an average of R$\sim$4500 for a 1” slit. SOXS also has imaging capabilities in the visible wavelength regime. It is designed and optimized to observe all kinds of transients and variable sources. The final destination of SOXS is the Nasmyth platform of the ESO NTT at La Silla, Chile. 
The SOXS consortium has a relatively large geographic spread, and therefore the Assembly Integration and Verification (AIV) of this medium-class instrument follows a modular approach. Each of the five main sub-systems of SOXS, namely the Common Path, the Calibration Unit, the Acquisition Camera, the UV-VIS Spectrograph, and the NIR Spectrograph, are undergoing (or undergone) internal alignment and testing in the respective consortium institutes. INAF-Osservatorio Astronomico di Padova delivers the Common Path sub-system, the backbone of the entire instrument. We report the Common Path internal alignment starting from the assembly of the individual components to the final testing of the optical quality, and the efficiency of the complete sub-system.

\end{abstract}


\section{INTRODUCTION}
\label{sec:intro}  
SOXS, the two-channel, single object, medium resolution spectrograph, is designed to observe transient events and variable sources\cite{Schipani16,Schipani18,Sanchez18,Aliverti18, Schipani20}. The SOXS instrument consists of five sub-systems, namely Common Path (CP), Calibration unit (CU), Acquisition Camera (AC), UV-VIS spectrograph, and the NIR spectrograph. CP is the backbone of the instrument\cite{Claudi18, Claudi22}. It has opto-mechanical interfaces to the other four sub-systems and the NTT telescope. During observation, CP receives the $F/11$ beam from the telescope, and the dichroic onboard splits the incoming beam, sending 350-850~$nm$ to the UV-VIS spectrograph and 800-2000~$nm$ to the NIR spectrograph. Both the spectrographs will receive an $F/6.5$ beam from the CP. For acquisition and imaging, the CP drives the light to the 3.5’ x 3.5’ Andor Camera. For wavelength and flux calibration purposes during the daytime, the CP can also direct the light from the CU to the spectrographs.

For the assembly, alignment, integration, and verification of the CP sub-system, we have exploited an opto-mechanical approach. We have used a portable Coordinate Measuring Machine (pCMM) to place the opto-mechanical components onto the CP bench. In addition to that, optical feedback using an on-axis laser source and a telescope simulator was used to fine-tune and validate the goodness of the alignment of the individual components. The $F/\#$ of the CP exiting beam to the spectrographs, their positions and tilts, and the on- and off-axis PSF optical quality were verified after the integration.

\subsection{The Common Path}\label{cp}
\begin{figure} [ht]
   \begin{center}
   \begin{tabular}{c} 
   \includegraphics[height=5.5cm]{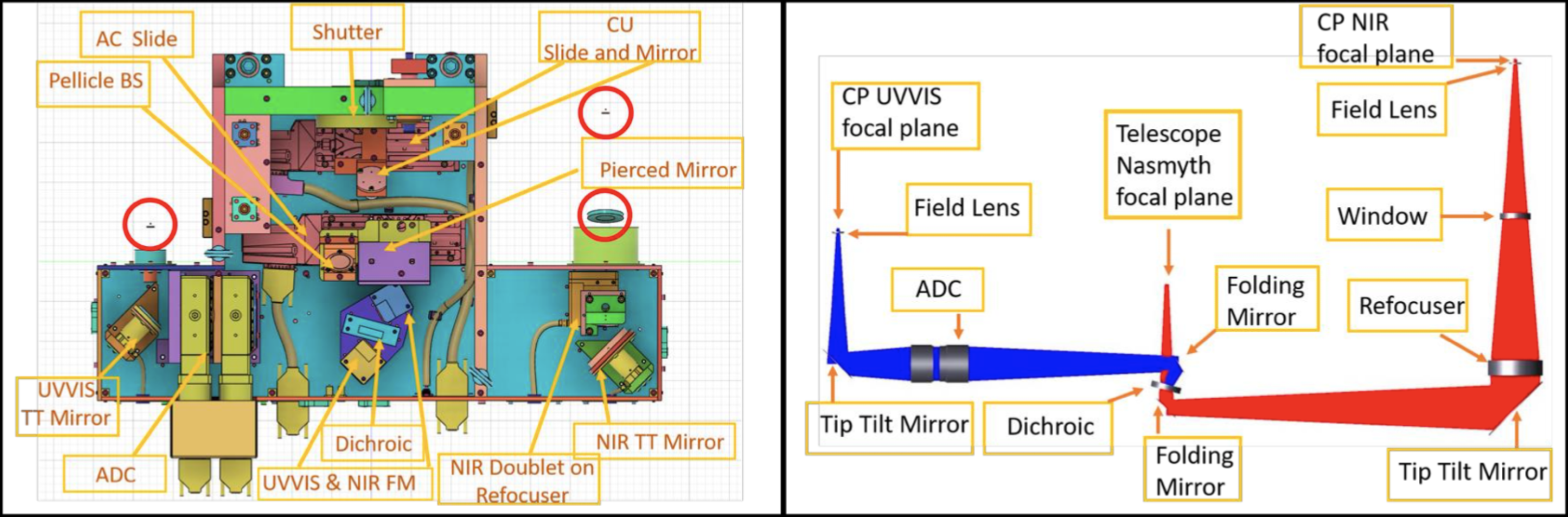}
   \end{tabular}
   \end{center}
   \caption[fig1] 
   { \label{fig1} 
\textit{Left panel}: Common Path CAD image displaying its components. \textit{Right panel}: The CP light path.}
   \end{figure}

The CP consists of a dichroic, 2 folding mirrors (FM), 2 tip-tilt (TT) mirrors (envisaged to compensate flexures), ADC in the UVVIS arm, refocuser mounted on a linear stage in the NIR arm, a linear stage to which the pierced mirror and pellicle are mounted (directing some or all light to the AC), a linear stage to which a folding mirror is mounted (directing the light from the CU towards the spectrographs), a PT100 temperature probe, and the instrument shutter. Figure~\ref{fig1} shows the CP, its components, and the light path.

The optical components UVVIS field lens, NIR window, and NIR field lens (marked within red circles in Figure~\ref{fig1} are formally a part of the CP, but physically present within the spectrographs. Without these components, the CP produces an $F/6.91$ beam at the UVVIS CP exit and an $F/6.8$ beam at the NIR CP exit.

\section{Alignment Strategy}

\begin{figure} [ht]
   \begin{center}
   \begin{tabular}{c} 
   \includegraphics[width=17cm]{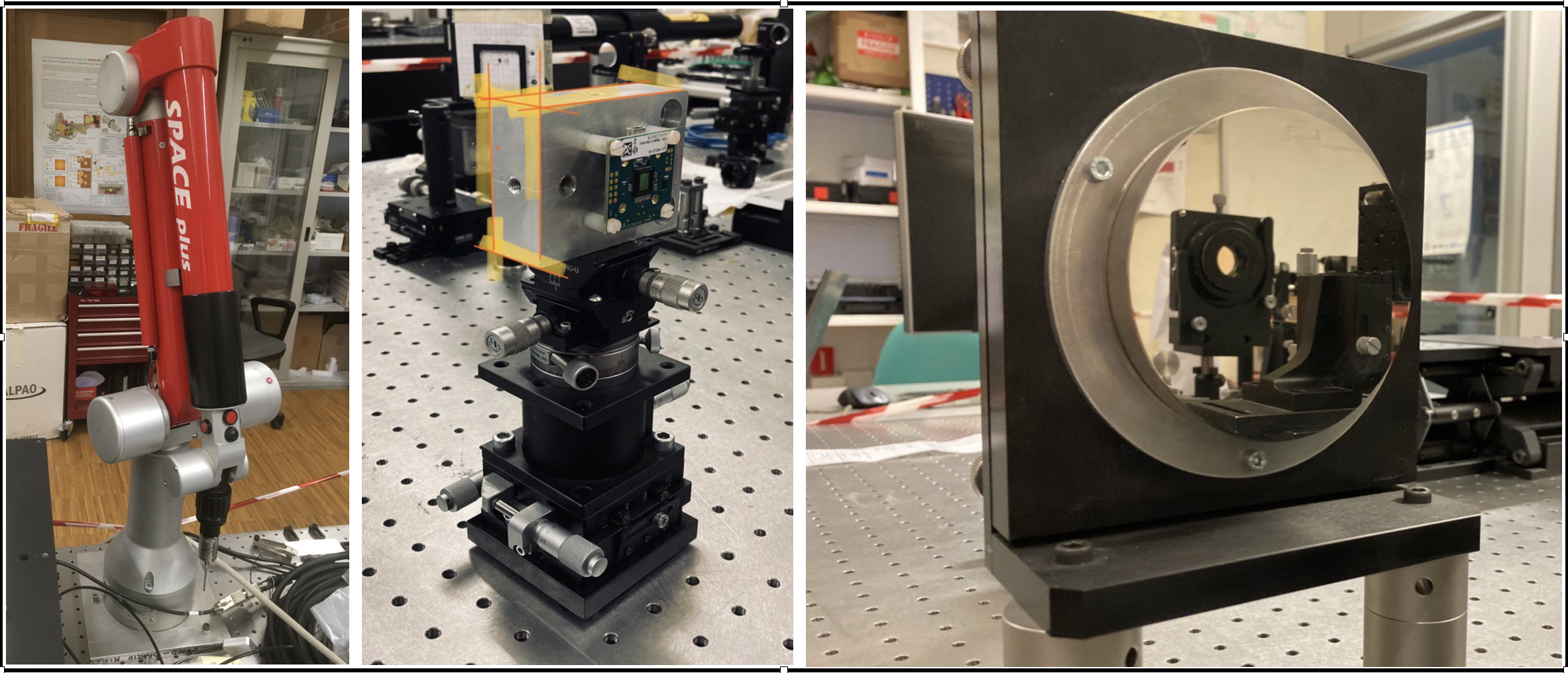}
   \end{tabular}
   \end{center}
   \caption[fig2] 
   { \label{fig2} 
\textit{Left}: A portable Coordinate Measuring Machine (pCMM). \textit{Middle}: Characterized CMOS detector mounted on XYZ-tip-tilt-rotation stage. \textit{Right}: $10~cm$ mirror used for alignment purposes.}
   \end{figure}

As mentioned earlier, the CP alignment follows an opto-mechanical approach. We used a pCMM to position the optical components (see Figure~\ref{fig2}), which has a measurement accuracy is about $30~\mu m$. We have therefore used large mirrors (e.g., $10~cm$ diameter) to minimize the error due to the pCMM whenever possible (see Figure~\ref{fig2}). 

\subsection{Characterized CMOS Detector}   
To accurately position the opto-mechanical components in decenter, in focus, and estimate the tilt of the beam after the component, we have characterized a CMOS detector mounted in a robust mount with 6 degrees of freedom (XYZ, tip, tilt, and rotation) as displayed in Figure~\ref{fig2}. 

The characterization was done using a converging beam coming from a 4D interferometer. We used the pCMM to have the mechanical measures. We were able to define a reference pixel and the distance from the intersection of three planes. This way we can position (inside the pCMM errors) the reference pixel of the CMOS detector on the nominal beam, checking for decenter and angles. 

\subsection{Alignment Set Up and Steps}
\begin{figure} [ht]
   \begin{center}
   \begin{tabular}{c} 
   \includegraphics[height=15cm]{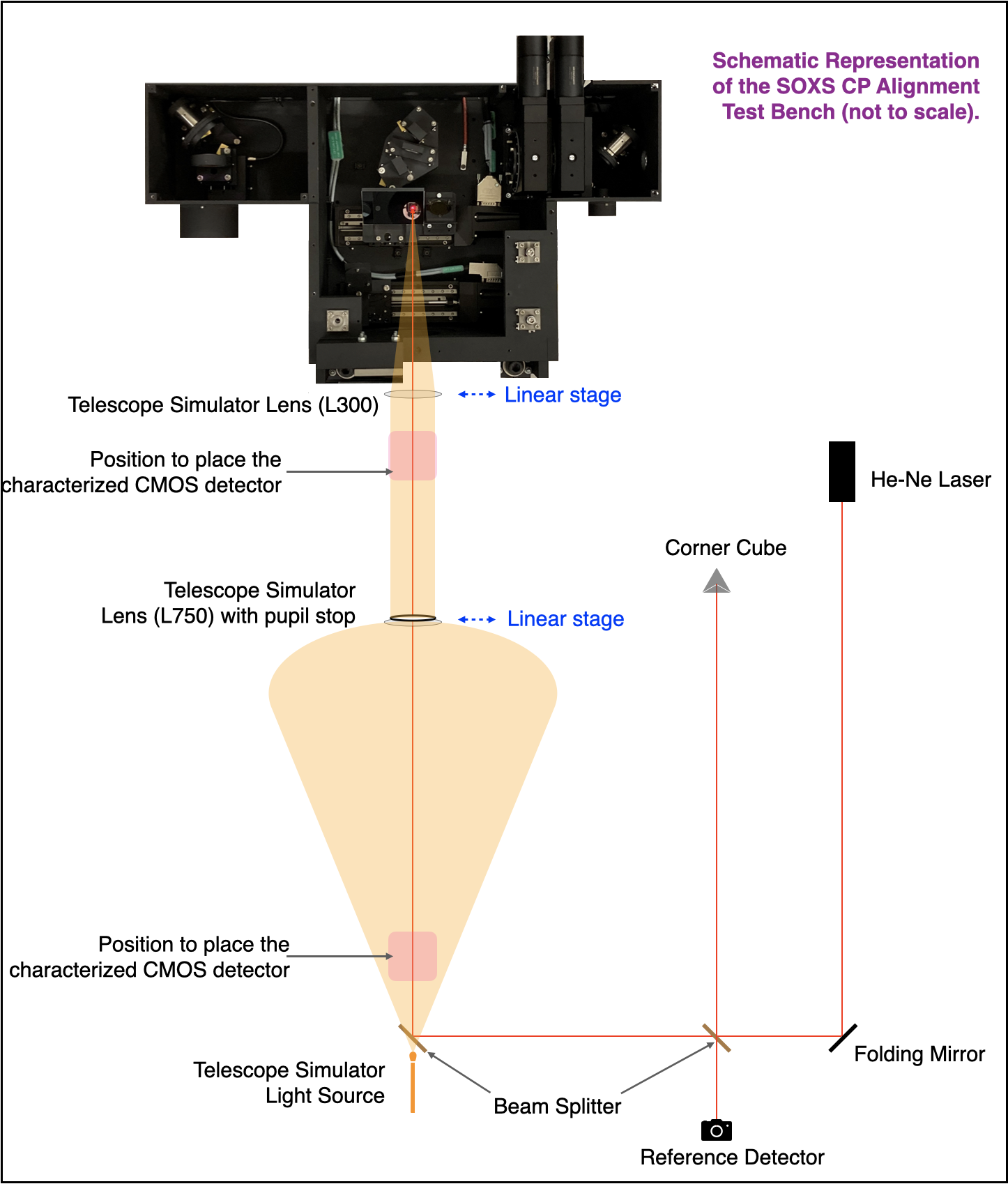}
   \end{tabular}
   \end{center}
   \caption[fig3] 
   { \label{fig3} 
Schematic representation of the CP alignment setup (not to scale).}
   \end{figure}

Schematic representation of the optical setup used to align the CP can be seen in Figure~\ref{fig3}. The CP is fixed to the optical bench using the kinematic mounts that will be used to mount the subsystem to the SOXS flange later on. The details of the kinematic mounts and actual image of the setup can be found in Biondi et al. paper\cite{Biondi20}. 

We could feed the CP either with a bright laser source or with an $F/11$ Telescope Simulator (TelSim). The details of the TelSim alignment is described in Biondi et al. article\cite{Biondi20}. The TelSim lenses ($L750$ and $L300$) were aligned to the nominal Laser beam with decenter $<40~\mu m$ and tip \& tilt $\sim$8” \& $\sim$4”, respectively. We achieved the TelSim PSF FWHM of $22.83~\mu m$, the nominal value being $22.62~\mu m$ and $F/\#$ of 10.94, nominal value being 11.00. The TelSim lenses can be moved out of the way to have the laser beam reaching directly the CP.

Shims were used to adjust the decenter, tip and tilt of the opto-mechanical components. Optical feedback was always used to confirm if the shims produced the right results.

The pCMM used the CP coordinate system for all its measurements. We used the CP-base, CP-side, and CP-front to create the CP coordinate system, with the origin matching the SOXS input focal position.

The alignment steps are enumerated below.
\begin{enumerate}
    \item The correct position of the dichroic with respect to the CP coordinate system was earlier determined using a CMM (which has $< 10~\mu m$ measurement precision). We used our pCMM to position the dichroic to the same location.
    \item Start with UVVIS arm or NIR arm.
    \item Align the FM. See Figure~\ref{fig4} for details of the procedure. We used the 10~$cm$ mirror, placing the mirror surface perpendicular to the nominal beam (using the pCMM) so as to send the reflected beam backwards eventually reaching the 'reference detector' (see Figure~\ref{fig3}). The tilt of the beam is verified also in this way.
    \item Align the TT mirror.See Figure~\ref{fig4} for details of the procedure. We used the 10~$cm$ mirror, placing the mirror surface perpendicular to the nominal beam (using the pCMM) so as to send the reflected beam backwards eventually reaching the 'reference detector' (see Figure~\ref{fig3}). The tilt of the beam is verified also in this way.
    \item Align the Refocuser / ADC. See Figure~\ref{fig4} for details of the procedure. Note that the ADC here should be internally aligned using a separate setup before placing onto the CP. Check Battaini et al.\cite{Battaini22} for the details of the ADC internal alignment and results.
    
\begin{figure} [ht]
   \begin{center}
   \begin{tabular}{c} 
   \includegraphics[height=11cm]{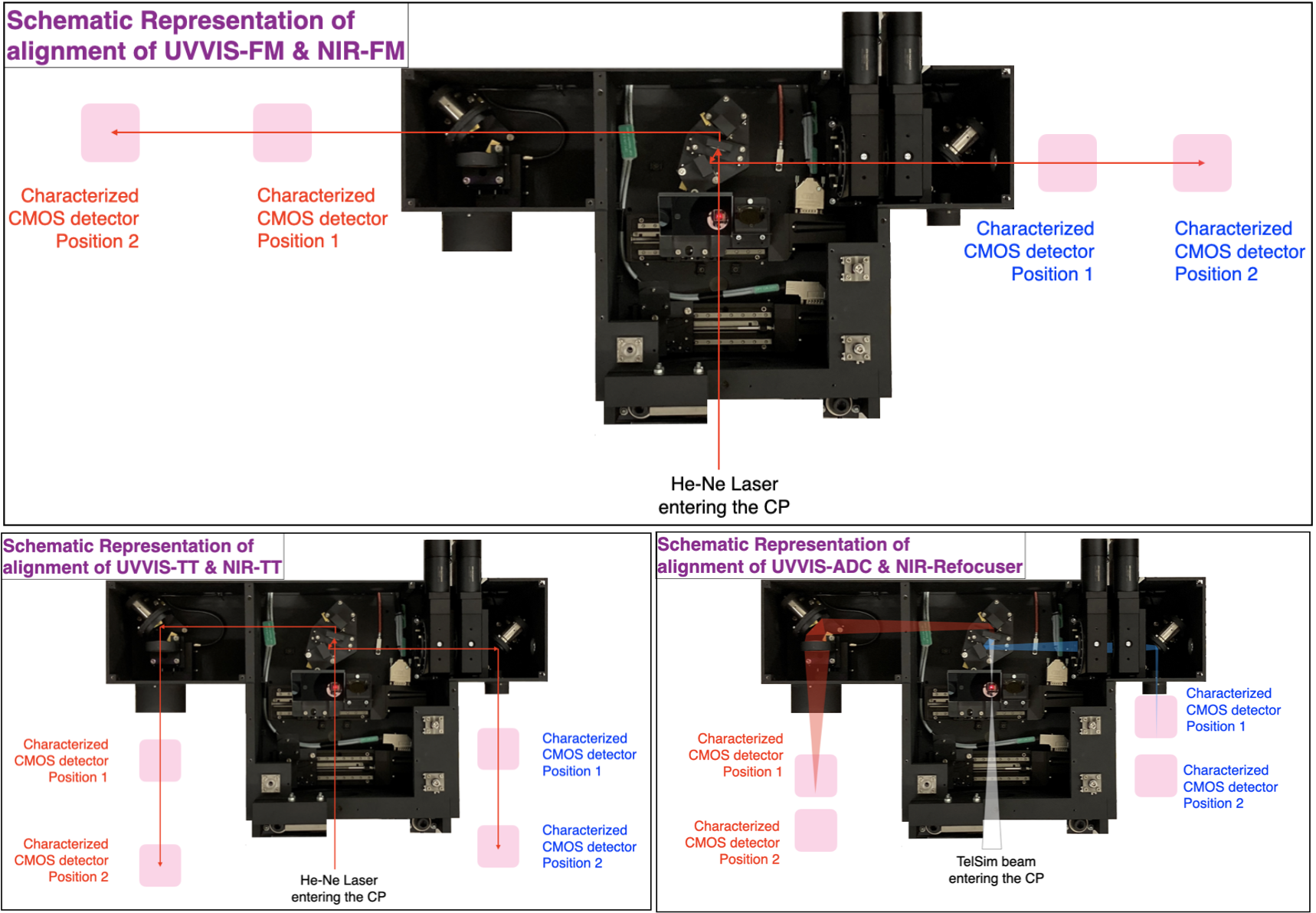}
   \end{tabular}
   \end{center}
   \caption[fig4] 
   { \label{fig4} 
\textit{Top}: Schematic representation of the alignment procedure of the CP-UVVIS-FM and CP-NIR-FM, \textit{Bottom Left}: Schematic representation of the alignment procedure of the CP-UVVIS-TT and CP-NIR-TT, \textit{Bottom Right}: Schematic representation of the alignment procedure of the CP-UVVIS-ADC and CP-NIR-Refocuser.}
\end{figure} 
    \item With the help of the pCMM and the characterized CMOS detector, check the PSF position, tilt of the exiting beam, PSF FWHM, exit $F/\#$, on- \& off-axis PSF quality, and the throughput for various narrow-band filters.
    \item Repeat the same for the other arm.
    \item The pierced mirror (which is a part of the AC, but physically present within the CP) and the CU folding mirror (which is a part of the CU, but physically present within the CP) are positioned using the pCMM. Measurements are taken to confirm that the mirror surfaces are at correct angles to the CP-base.
    \item The instrument shutter is installed and its functionality is checked.
    \item The PT100 temperature probe is installed onto the CP bench. Its functionality is verified using the instrument software.
    \item Cabling of all the wires within the CP is performed.
    \item All the linear stages with the final load is tuned to minimize the vibrations and named locations are found and registered at the configuration files for the instrument software.
    
\end{enumerate}

\section{Alignment Verification}   
After the alignment and integration of the individual opto-mechanical components, the verification of the subsystem is performed. These verification included the following tests.
\begin{enumerate}
    \item Decenter, focus and tilt of the exiting beam: The characterized CMOS detector is positioned so that the reference pixel is at the nominal, on-axis, focus position of the CP UVVIS/NIR exit focal position. At this point, the decenter is measured using the actual position of the PSF. The tilt is measured by taking multiple images at different optical axis positions and compensating for the tilt of the linear stage measured by the pCMM.
\begin{table}[ht]
\caption{CP UVVIS and NIR exit beam results. Note that the nominal $F/\#$ values are mentioned in Section~\ref{cp}. For the reported values here, we assumed a conservative estimation on uncertainty for the tilt because of the measuring conditions and technique.} 
\label{table1}
\begin{tabular}{|l|l|l|l|l|l|l|} 
\hline
\rule[-1ex]{0pt}{3.5ex}  & Decenter–X ($\mu m$) & Decenter–Y ($\mu m$) & Decenter–Z ($\mu m$) & Tilt-X (") & Tilt-Y (") & $F/\#$ \\
\hline
\rule[-1ex]{0pt}{3.5ex}  UVVIS exit beam & 230 $\pm$ 60 & -145 $\pm$ 60 & $<$30 $\pm$ 60 & -413 $\pm$ 100 & 37 $\pm$ 100 & 6.93 \\
\hline
\rule[-1ex]{0pt}{3.5ex}  NIR exit beam & 45 $\pm$ 60 & 3 $\pm$ 60 & $<$30 $\pm$ 60 & -250 $\pm$ 32 & -255 $\pm$ 32 & 6.94 \\
\hline 
\end{tabular}
\end{table}

\begin{figure} [ht]
   \begin{center}
   \begin{tabular}{c} 
   \includegraphics[width=17cm]{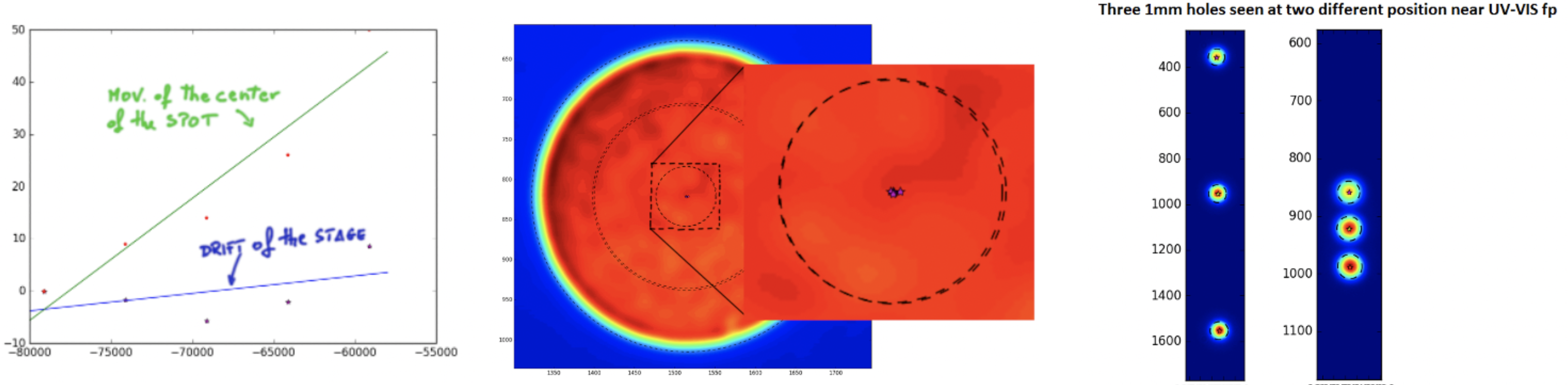}
   \end{tabular}
   \end{center}
   \caption[fig5] 
   { \label{fig5} 
\textit{Left}: Plot showing the tilt of the UVVIS PSF and that of the linear stage, \textit{Middle}: Finding the center of the PSF image, \textit{Right}: $F/\#$ measurement images for the UVVIS side.}
\end{figure} 
    
    \item $F/\#$ of the exiting beam: A mask with three holes of 1~$mm$ diameter is positioned in the collimated beam (between the $L750$ and $L300$ lenses of the TelSim). Keeping a detector at two different locations of the exiting beam, and knowing the distance between the two positions, the $F/\#$ of the exiting beam is estimated.
    \item On- and off-axis PSF quality: Using the characterized CMOS detector, the PSF of the TelSim is taken at the CP UVVIS/NIR exit focal position. For the off-axis positions, the fiber source of the TelSim is moved by known amount. The numbers obtained are compared with ZEMAX simulated values.
\begin{figure} [ht]
   \begin{center}
   \begin{tabular}{c} 
   \includegraphics[width=12cm]{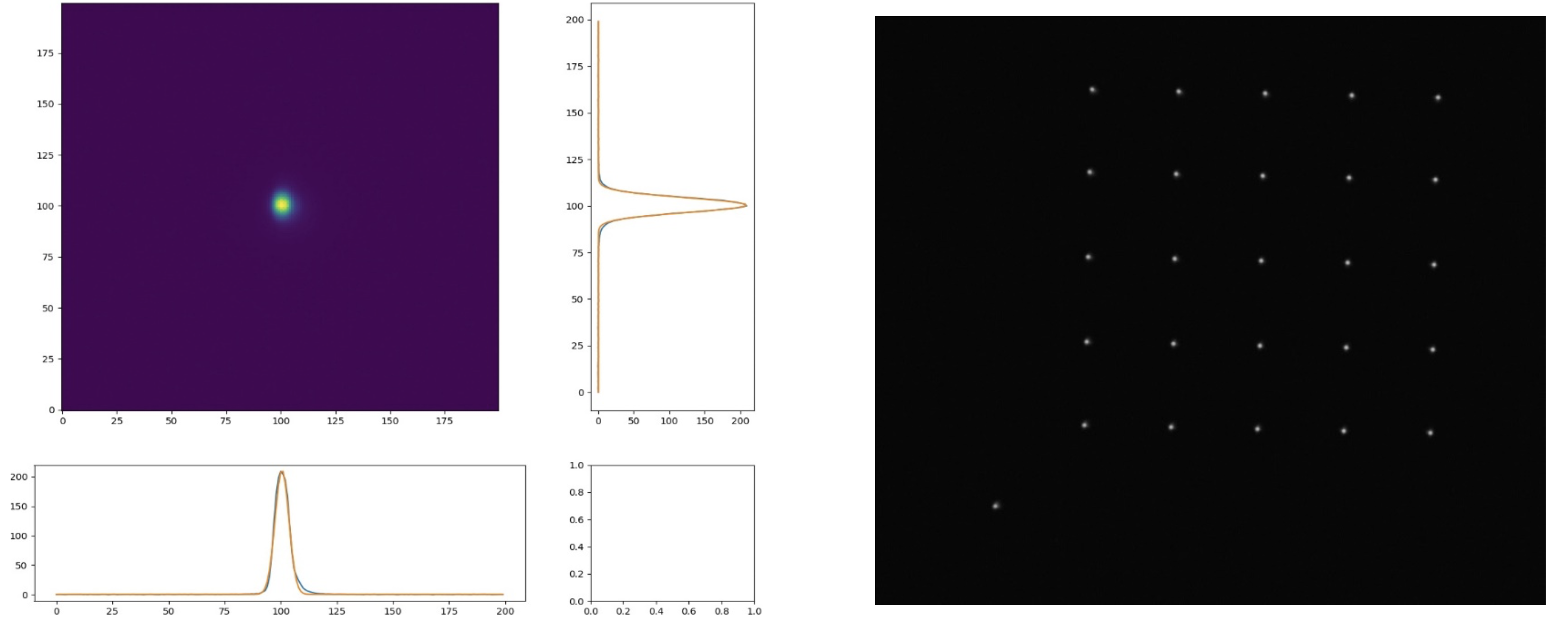}
   \end{tabular}
   \end{center}
   \caption[fig6] 
   { \label{fig6} 
\textit{Left}: UVVIS PSF, \textit{Right}: UVVIS on- and off-axis PSFs.}
\end{figure} 
    
    \item Transmission efficiency: Using flexOptometer, which is a multi-channel radiometer, the optical efficiency of the CP is estimated for both the UVVIS and NIR arms. The input (or reference) value is taken at the SOXS focal position (same location as that of the telescope focus as depicted in Figure~\ref{fig1}) and the output measurement is taken the CP UVVIS/NIR focal position. A white light source and various narrow-band filters are used for estimating the optical efficiency at different wavelengths.
\end{enumerate}

\begin{figure} [ht]
   \begin{center}
   \begin{tabular}{c} 
   \includegraphics[width=10cm]{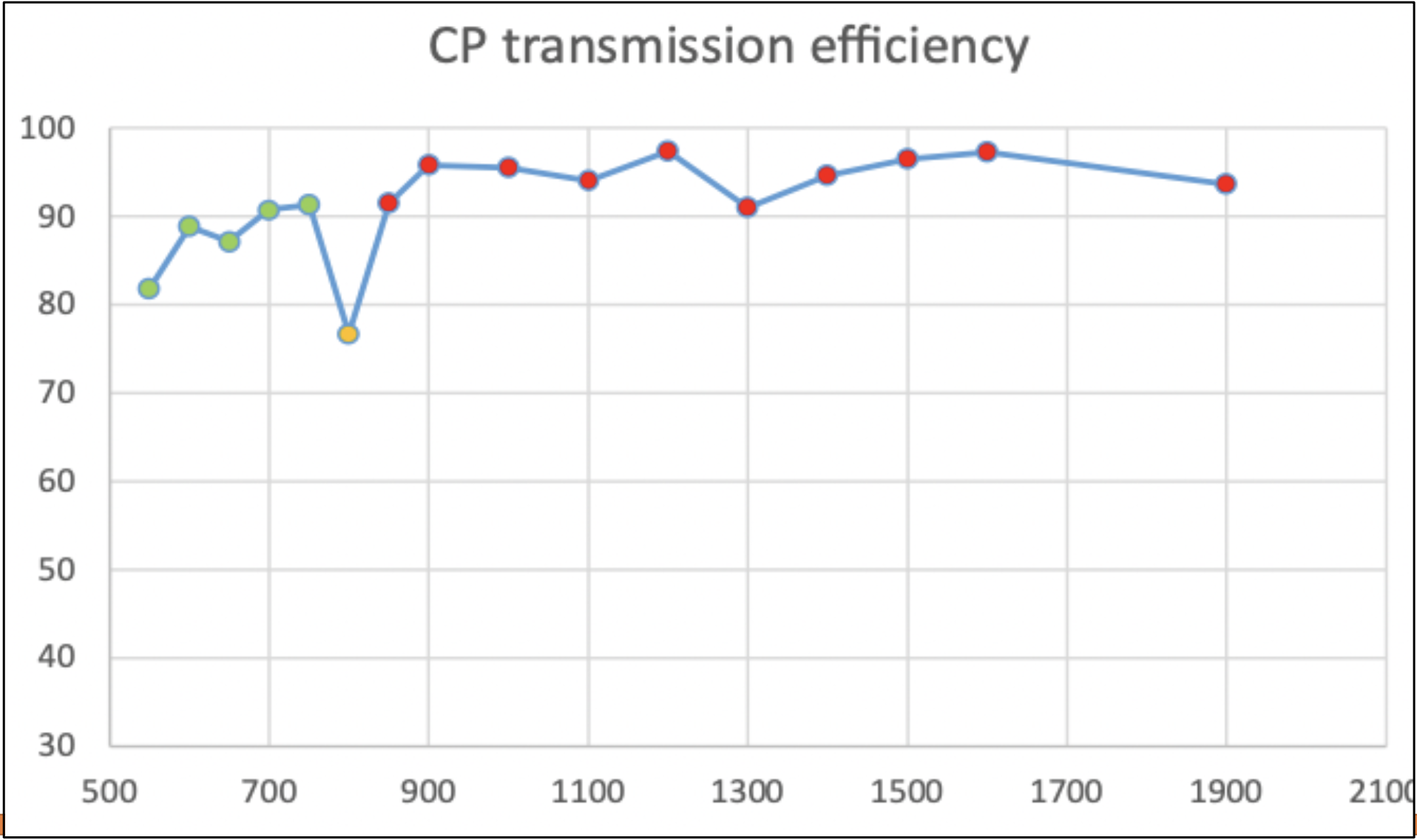}
   \end{tabular}
   \end{center}
   \caption[fig7] 
   { \label{fig7} 
Transmission is always above 80\% within the operational limits our instruments for the wavelength range from 550~$nm$ to 1900~$nm$. For the measurement, we have used a set of narrow-band filters of 10~$nm$ width. For 800~$nm$ wavelength, the efficiency is 76.7\%. This measurement comes only from the UVVIS side. The dichroic transmission is from 800-850~$nm$.

}
\end{figure}

The results are tabulated in Table \ref{table1} and displayed in Figure~\ref{fig5},~\ref{fig6}, and ~\ref{fig7}.

\section{Conclusion}
 
The SOXS Common Path is aligned, integrated and it passed all the verification tests. In addition to the optical verification, it has also passed the internal software and electrical tests with the instrument software and final electrical configurations respectively. 

\begin{figure} [ht]
   \begin{center}
   \begin{tabular}{c}
   \includegraphics[width=10cm]{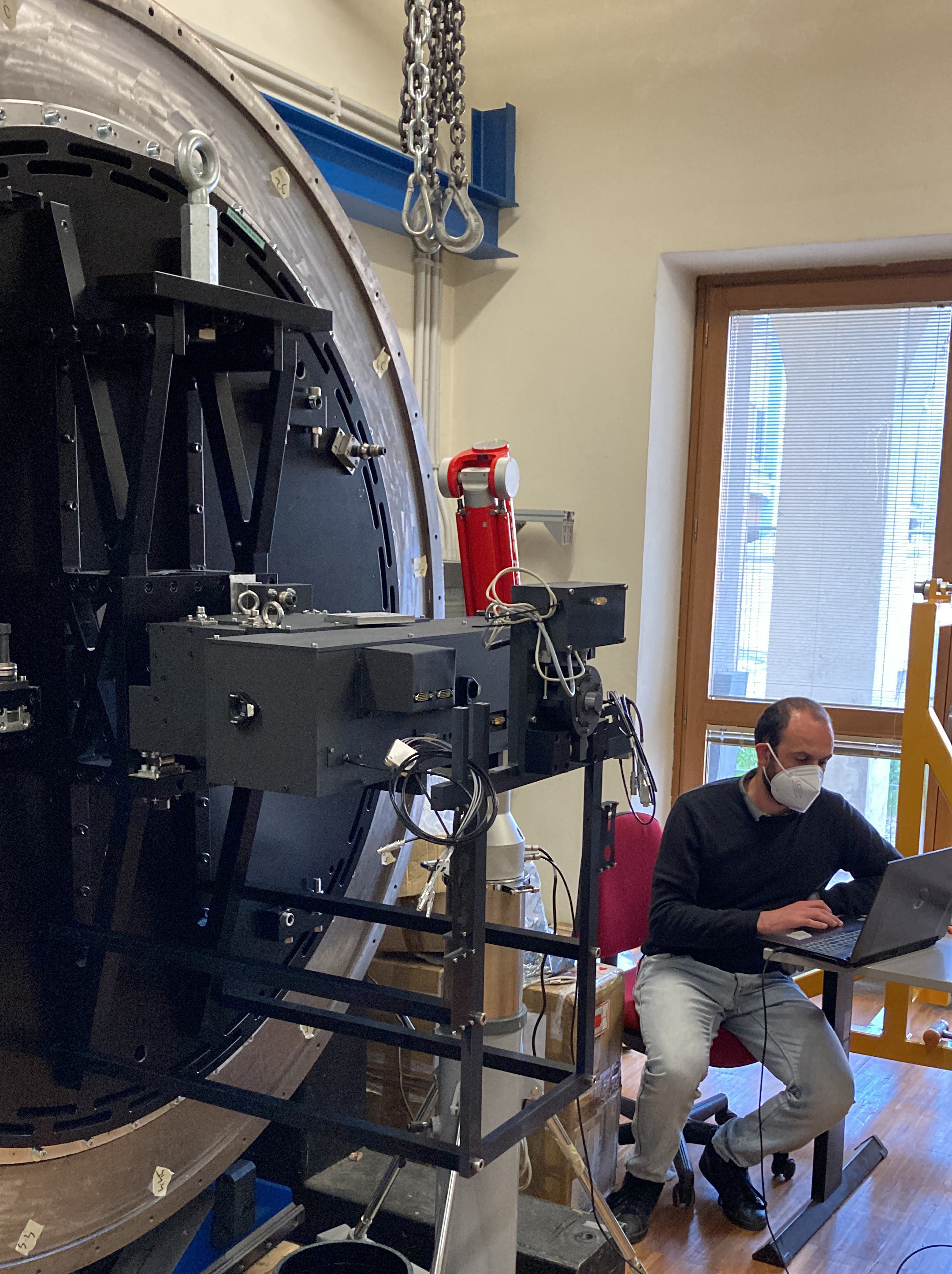}
   \end{tabular}
   \end{center}
   \caption[fig8] 
   { \label{fig8} 
SOXS Common Path mounted onto the SOXS flange at the INAF-Padova laboratories. The kinematic mount stability between the CP and the flange is also verified and found to be well within requirements.}
\end{figure} 

The sub-system is completed. We also verified the mechanical interfaces (using the kinematic mounts) with the SOXS flange. Figure~\ref{fig8} shows the common path mounted onto the SOXS flange. For more details about the integration of the CP (and other sub-systems) to the SOXS flange, refer to Aliverti et al. article\cite{Aliverti22}.

\acknowledgments 

KKRS, FB, and RC express their gratitude to the colleagues, INAF-PD administrative staff, and SOXS team for the continued support which facilitated the smooth AIVT process of the SOXS CP in spite of the COVID-19 pandemic in the last two years.

\bibliography{main} 
\bibliographystyle{spiebib} 

\end{document}